\documentstyle[aas2pp4]{article}
\lefthead{Ruffolo}
\righthead{Charge States of Solar Cosmic Rays}

\begin{document}

\title{Charge States of Solar Cosmic Rays\\ 
   and Constraints on Acceleration Times and Coronal Transport}

\author{D. Ruffolo}
\affil{Department of Physics, Chulalongkorn University, Bangkok
10330, Thailand}

\begin{abstract}
     We examine effects on the charge states of energetic ions associated
with gradual solar flares due to shock heating and stripping at high
ion velocities.  Recent measurements of the mean charges of various
elements after the flares of 1992 Oct 30 and 1992 Nov 2 allow one to
place limits on the product of the electron density times
the acceleration or coronal residence time.  In particular, any 
residence in coronal loops must be for $<$0.03 s, which rules out
models of coronal transport in loops, such as the bird cage model.
The results do not contradict models of shock acceleration
of energetic ions from coronal plasma at various solar longitudes.
\end{abstract}


\section{Introduction}

Although particles from solar flares have been observed for 50
years (\cite{f46}), there are still many unresolved questions about
their acceleration sites and mechanisms.  Much recent research has
focused on a simple classification scheme proposed by Pallavicini,
Serio, \& Vaiana (1977).  In modern terminology, ``impulsive'' solar
flares are typically defined as those with a short ($\lesssim 1$ h)
duration of X-ray emission, while ``gradual'' flares have a longer
X-ray duration.  These two classes of flares have been found to have
several distinguishing characteristics.  For example, X-ray emission
from impulsive flares is observed to come from compact regions
at low coronal heights ($\lesssim 10^4$ km),
while X-rays from gradual flares tend to arise from broader regions or
from large coronal loops up to $\sim10^5$ km above the photosphere.
In addition, it is widely believed that the
main acceleration mechanism for particles escaping from impulsive flares is
stochastic acceleration (second-order Fermi acceleration by
gyroresonant plasma waves; e.g., \cite{tr92,mv93}),
while gradual flares are associated with shock
acceleration higher in the corona (e.g., \cite{cea86,lr86}), though not all
observations support this paradigm (e.g., \cite{mea92}).

A more controversial issue is why solar cosmic rays are observed at 
locations that are magnetically connected to solar longitudes far from 
the longitude of the flare site.  Because interplanetary diffusion is
highly anisotropic, inhibiting motion perpendicular to the magnetic
field (e.g., \cite{p82}), for decades it was assumed that particles
are transported to other solar longitudes within the solar corona,
whence they escape to travel along the interplanetary magnetic
field to the observer.  The seminal quantitative model of Reid (1964)
assumes isotropic, two-dimensional, coronal diffusion, and
gives specific predictions for the injection rate of particles
into the interplanetary medium as a function of time and coronal
distance from the flare site.  It is not clear
exactly what mechanisms give rise to coronal diffusion, though
Newkirk \& Wentzel (1978) presented the ``bird cage''
model of rigidity-independent coronal transport, in which 
flare-accelerated particles bounce back and forth 
inside coronal loops and occasionally transfer to other loops at the 
footpoints.  On the other hand, it has recently
been proposed that for gradual flares, with are often associated with
large, interplanetary shocks, particles are freshly accelerated on
open field lines at different heliolongitudes (\cite{mgh84,r90}).  
It is further assumed that coronal transport does not occur, and that for 
impulsive flares, which are usually not associated with interplanetary 
shocks, the narrower longitudinal dispersion (\cite{rea90}) 
is due to the spreading of magnetic field lines in the solar corona or the 
interplanetary medium.  Therefore, it is no longer clear whether several 
decades worth of solar cosmic ray observations were providing information 
about the solar flare site, as previously assumed, or about acceleration at 
coronal or interplanetary shocks.  Further progress in the interpretation of
the information embodied in solar cosmic ray observations thus
depends on the resolution of this issue.

Recent observations which can shed much light on this issue concern
the charge state distributions of ions accelerated as a result of solar 
flares.  Such distributions for various elements provide a
rich source of information on the conditions of particle
acceleration and escape from the corona.  Results of the ULEZEQ instrument on 
board the {\it ISEE-3} spacecraft indicated that for several gradual flares, 
the charge states of 9 different elements were not consistent with the same 
temperature (\cite{lea84}), but rather had apparent ionization temperatures 
of 1 to 8$\times10^6$ K.  In contrast, summed results for several less 
powerful impulsive flares suggested that charge states of Si and Fe were
characteristic of significantly hotter plasma, as might be expected
if particle acceleration takes place at the site of a compact,
impulsive flare (\cite{lea87}).  More recently, measurements by three 
instruments on board the {\it SAMPEX} mission (\cite{oea95,lea95,mea95}) 
have generally confirmed the earlier results for gradual flares, and also 
provide measurements for more elements, with better statistical accuracy, and 
for a broader energy range.  A possible explanation for why the mean charges 
of energetic ions from gradual events are not consistent with a single 
temperature was given by Mullan and Waldron (1986), who proposed that 
photoionization by flare X-rays changes the ionization equilibrium in the 
plasma from which ions are accelerated, and obtained good quantitative 
agreement with the apparent ionization temperatures for various elements.  
Another possible explanation in terms of freeze-out temperatures was 
discussed by Mason et al.\ (1995).

In this report, we examine effects on the charge states
of energetic ions associated with gradual solar flares
due two processes which would be expected to occur at the acceleration site.
Although the physical conditions of the ion acceleration are
poorly known, it is possible to roughly estimate these effects.
If we accept that ions escaping after gradual flares
arise from shock acceleration high in the corona, then one might expect
significant electron heating, which is a general phenomenon
of shocks in astrophysical plasmas (\cite{sea88}).
Another effect is that as the ions' velocity increases, their
electrons are more likely to be stripped off (\cite{hea84,lea95}).
We point out that if the ions spend a
sufficiently long time at the acceleration site or in the solar corona, then 
these effects would influence the charge
states in a manner that is inconsistent with the observations.
This places limits on the acceleration time, and on the time spent in
the solar corona after acceleration.  These limits in turn rule out 
models of ion acceleration or coronal transport (such as the bird cage
model) which require that escaping ions spend over $\sim0.03$ s in a
magnetic loop, while they do not contradict models in which
ions promptly escape after acceleration on open field lines.

\section{Effect of Shock Heating and Ion Velocities on Mean Charge States}

As mentioned above, there are several recent reports of
high-quality data on the mean charges of solar
cosmic ray ions.  Here we will mainly consider the results of Leske et
al.\ (1995), who measured the mean charges of 12 elements with a
high statistical accuracy at the relatively high energy of $\sim$15-70 MeV
nucleon$^{-1}$.  Very similar results were obtained for two large,
gradual flares on 1992 Oct 30 (X-ray magnitude X1, importance 2B, at
22$^\circ$S, 61$^\circ$W) and 1992 Nov 2 (X9, 2B, at 23$^\circ$S,
90$^\circ$W; \cite{sgd93}).

To place conservative limits on the acceleration or coronal residence 
time for solar cosmic rays, we assume that the charge states are
originally characteristic of a relatively cool coronal plasma 
in thermal equilibrium, and that other effects, such as shock heating or 
increased stripping for faster ions, would tend to increase the
charge.  The lowest apparent ionization temperature that was
accurately measured by Leske et al.\ (1995) is that for Si, 
1.75$\pm$0.11$\times10^6$ K, which was essentially the same on both 1992 
Oct 30 and 1992 Nov 2.  For a conservative limit, we use an upstream 
plasma temperature of $T_u=1.5\times10^6$ K,
which is also the average value obtained by Mullan \& Waldron (1986)
for their explanation in terms of X-ray photoionization.

The first effect which we consider is that of electron heating by
shocks propagating through the corona, since it is widely believed
that cosmic rays associated with gradual flares are accelerated by such 
shocks.  Observations of a large number of fast-mode shock
crossings of interplanetary shocks and bow shocks of Earth, Jupiter,
and Uranus (\cite{tea87,sea88}) imply that a fixed fraction of
0.12$\pm$0.05 of the incident ion ram kinetic energy goes into
electron heating, i.e., into increasing the electron enthalpy.
From Figure 3 of Schwarz et al.\ (1988), we see that the 
jump in the electron temperature, $\Delta T_e$, 
across the shocks in their sample is typically given by 
\begin{equation}
%\frac{\Delta T_e}{\mbox{K}} = (3\mbox{ to }9)
%\times\frac{V_u^2-V_d^2}{\mbox{km}^2\mbox{ s}^{-2}},
\Delta T_e = (3\mbox{ to }9 \mbox{K})\times 
  \frac{V_u^2-V_d^2}{\mbox{km}^2\mbox{ s}^{-2}},
\end{equation}
especially for $V_u^2-V_d^2>2.5\times10^5$ km$^2$ s$^{-2}$, where
$V_u$ and $V_d$ are the bulk flow speeds upstream and downstream,
respectively, in the shock (de Hoffmann-Teller) frame.

To estimate the shock speed, we note that magnetohydrodynamic
simulations indicate that the initial coronal mass ejection
(CME) speed is roughly twice the 
average propagation speed of the forward shock to the Earth
(\cite{hea95}).  Based on the $K_p$ index of geomagnetic activity
(U.S.\ National Geophysical Data Center, via World Wide Web), we
estimate the initial CME speed for both flare events to be $V_u\approx
1500$ km s$^{-1}$.  Assuming $V_d=V_u/4$ for a strong shock,
the downstream electron temperature, $T_d$, is
estimated to be 6 to 19$\times10^6$ K.  Note that since shock speeds
determined from Type II radio bursts can be higher than initial CME
speeds (e.g., \cite{hea95}), using such speeds would yield higher
downstream temperatures.  A conservative limit on $nt$ is obtained by
considering $T_d=7\times10^6$ K, which was also invoked in a 
mechanism proposed by Luhn et al. (1984) in which
ions are suddenly heated to this temperature and the different
apparent ionization temperatures for various elements are ascribed to
different rates of approach to the new ionization equilibrium.  
For comparison, we also consider the effect of a higher value,
$T_d=1.5\times10^7$ K.

To calculate charge state distributions, we used the ionization and 
recombination rate coefficients of Shull \& Van Steenberg (1982), correcting 
the misprints in those tables reported in the errata (ApJS, 49, 351) and
by Arnaud \& Rothenflug (1985).  The rate of change of $n_q$, the
number density of ions of charge $q$, is given by
\begin{equation}
\frac{dn_q}{dt} = n_e[I_{q-1}n_{q-1}-(I_q+R_{q-1})n_q+R_qn_{q+1}],
\end{equation}
where $I_q$ and $R_q$ are temperature-dependent
rate coefficients (in cm$^3$ s$^{-1}$) of ionization from charge $q$
to $q+1$ and recombination from $q+1$ to $q$, respectively.
For ten elements, and different $nt$ values, we used Euler's method to
calculate the mean charges expected due to a) shock acceleration, in which 
particles are sometimes at $T_u$ and sometimes at $T_d$ (we used equal times 
for a conservative limit), and b) the mechanism of Luhn et
al.\ (1984), in which ions are immersed completely in a new $T_d$.
Table 1 shows the mean charges of various elements for case a).
Note that for a higher $T_d$, smaller values of $nt$ yield similar
results.  For case b), almost identical mean charges were
obtained for $nt$ values half as large.
It is clear that neither mechanism is able to explain the observations.
In particular, at $nt=2\times10^{10}$ cm$^{-3}$ s, the mean charges
calculated for Si, S, Ar, Ca, and Ni were all above the $1\sigma$ limits of
Leske et al.\ (1995), while those for O, Ne, and Fe were too low.
We therefore place a conservative limit of $nt<2\times10^{10}$ based on shock
heating, where $t$ is the total residence time at the shock.

Next we consider the effect of the ion velocity.  The new observations of
solar cosmic-ray charge states after gradual flares at higher
energies, $\gtrsim10$ MeV (\cite{lea95,oea95,tea95}), 
are in general agreement with those at lower energies
(\cite{lea84,mea95}; note, however, that different mean charges are
obtained for Fe, as discussed in the latter reference), 
which implies that little or no additional stripping occurs during the 
acceleration to higher energies or escape from the corona (\cite{lea95}). 
Firstly, let us consider a limit on the integrated electron density
times residence time in the corona after ions have been accelerated
to their final energy, assuming that all changes in charge states
relative to equilibrium at the plasma temperature occur during 
post-acceleration motion through the corona.  Ions of 15 MeV nucleon$^{-1}$
(at the lower end of the energy range measured by \cite{lea95})
move at $\approx 6.4$ times the rms thermal speed of electrons at our
assumed plasma temperature of $T_u=1.5\times10^6$ K.  At these relative
speeds, collisional ionization is far above threshold, and varies
slowly with the relative speed, so we can neglect the effect of the
thermal spread and approximate the ionization cross sections
by assuming $v_{\mbox{\scriptsize rel}}=v_{\mbox{\scriptsize ion}}$.
We use equation (2) to calculate the charge state distributions,
but neglect the recombination rates because the 
charge states characteristic of $T_u$ are far from the equilibrium for
such a high velocity.  For ionization rates, we employ
the energy-dependent ionization cross sections $\sigma_q$
of Arnaud and Rothenflug (1985), and set 
$I_q=\sigma_qv_{\mbox{\scriptsize rel}}$.  Again, the charge state 
observations for various elements cannot be explained for any value of $nt$.  
As for the case of shock heating, the tightest limits on $nt$ are
obtained for Ar, Ca, S, and Si.  The $nt$ values for which the calculated 
mean charges exceed the observed mean charges by one standard deviation are 
shown in the third column of Table 2.  From these we 
conclude that $nt$ should not exceed about 3$\times10^9$ cm$^{-3}$ s 
during transport through the corona.

Similarly, if we assume that all deviations from thermal-equilibrium
charge states occur during acceleration, we can derive limits on the
acceleration time.  Here we need to consider the changing velocity of the 
ion.  In fact, for slower ions the rate of ionization is faster, though the
ultimate equilibrium charges would be lower.  To be conservative, we only 
consider the final $e$-folding in momentum from 61 to 167 MeV $c^{-1}$, 
corresponding to kinetic energies from 2 to 15 MeV, and 
assume as a rough approximation that the rate of change of $\log(p)$
is constant (which requires that the spatial diffusion coefficient 
should be roughly constant over that energy range).
The resulting $nt$ values for which calculated mean charges exceed the
observed mean charges by one standard deviation are shown in the second 
column of Table 2, where $t$ represents the momentum $e$-folding time.
We can conclude that the maximum momentum $e$-folding time during 
acceleration, which is commonly called the acceleration time, is
given by $nt<2\times10^9$ cm$^{-3}$ s.
Since the limits we derive for acceleration and coronal residence
times are of a similar magnitude, we can alternatively say that
$nt<3\times10^9$ cm$^{-3}$, where $t$ is
the acceleration time plus the residence time in the corona.
The earlier limit derived from the effect of shock heating 
is less stringent than that from the effect of the ion velocity, 
so we will use the latter in the following discussion.

\section{Discussion}

We have implicitly made the standard assumption that the charge
states of solar cosmic rays are fixed in the corona and unaffected by
interplanetary transport.  This is justified by considering the above
limits.  Given that solar wind densities are on the order of 4-12
cm$^{-3}$ (\cite{f90}) near Earth, and vary as the inverse square of
the radius from the Sun, ions could travel for years throughout the inner
solar system without being substantially stripped.
However, for the much higher densities within the solar corona, 
these limits on $nt$ correspond to much shorter times.

A related point is that Tylka et al.\ (1995) have stressed that most
measurements of the mean charge of iron are incompatible with acceleration
directly from the ambient solar wind (Reames 1990).
This point is supported by previous theoretical and observational studies 
(e.g., \cite{ts89,tea90}), which found that at energies $\gtrsim1$ MeV,
interplanetary shocks only re-accelerate an existing population of
somewhat lower energy.  The observed charge states seem to
require an initial acceleration from coronal material.  
This does not rule out the possibility that a CME-driven shock 
accelerates particles while inside the corona, or perhaps 
near the Sun (\cite{krs90}) while pushing coronal material in front of it
(\cite{bta96}).
                       
We consider the implications of limits on
the product of the electron density and the time of acceleration 
or residence in the corona for models of coronal transport.
While the flare of 1992 Oct 30 was at a solar longitude of 61$^\circ$ W
that could be magnetically well-connected to the detectors near the Earth, the
flare of 1992 Nov 2 was at $\approx90^\circ$ W, so the hypothesis
of coronal transport would have to account for a net displacement of
about $4\times10^5$ km.  If we consider the limit of $nt<3\times 10^9$
cm$^{-3}$ s in the corona after and during acceleration,
the electron densities in various regions of the corona (and the
corresponding limits on $t$) are roughly $10^{11-12}$ cm$^{-3}$ (0.003-0.03 
s) in coronal loops, and $5\times10^7$ cm$^{-3}$ (60 s) at a height of $10^5$ 
km in the quiet Sun (Foukal 1990).  
As field lines outside the loops rapidly open up with height, the
density rapidly decreases, until $n\sim10^5$ cm$^{-3}$
($t\lesssim3\times10^4$ s) at $r=2 R_{\odot}$.

These limits essentially rule out the ``bird cage'' mod\-el of
rigidity-independent coronal transport (New\-kirk \& Wentzel 1978) for 
escaping ions from these gradual flares.  In this model, particles are stored
inside coronal loops, mirroring back and forth between the footpoints and 
occasionally transferring to a different loop.  However, the above limits 
show that the ions which escape could have traveled at most 1.6$\times10^3$ 
km inside a coronal loop, which would only be a small fraction of the
loop's length, and is far shorter than the coronal distance from the
site of the 1992 Nov 2 flare to the footpoint of the detectors' magnetic 
field lines.  Thus if the ions which are observed are accelerated inside
a loop, the acceleration time must be very fast ($<0.02$ s), and
the ions must escape immediately and cannot be transported along
the loop.

If a mechanism for coronal transport does not involve the motion
of ions in coronal loops, e.g., by transport along current sheets
(\cite{fs72}; note that such transport would be rigidity-dependent),
the above limit on the time spent at a coronal height of $\sim10^5$ km
allows a 15 MeV nucleon$^{-1}$ ion to travel about 3$\times10^6$ km.  
However, the above limits on $nt$ are very conservative, in that they permit 
an ion's mean charge to change from the value characteristic of
$1.5\times10^6$ K to the measured values, while in fact there seems
to be little change in the mean charges of observations spanning two
orders of magnitude in energy.  It is possible that {\it SAMPEX} observations
could yield $d\langle Q\rangle/dE$ for a single instrument, placing 
more severe constraints on $nt$ and on coronal transport models in general
(see also \cite{oea97}).

Our results favor models in which the acceleration of ions
observed after gradual flares takes place on open magnetic field lines 
(e.g., \cite{lr86}), and in which the longitudinal distribution of solar 
cosmic rays from such events is due to acceleration distributed over a range 
of heliolongitudes.  

Since the effects of shock heating and ion velocities cannot explain
the different apparent ionization temperatures for various elements,
the best quantitative explanation is that of Mullan and Waldron (1986).
An requirement of this model is that similar X-ray fluxes must be present at 
the acceleration site for all of the observed flares, over 
a time sufficient for ionization equilibrium to be established,
before a shock arrives and accelerates the ions.  Note that the X-ray flux
should decrease with increasing coronal distance, and be eclipsed for
low heights at large coronal distances.  If one believes
that observed ions are accelerated near the footpoint of the observer's
field line, then the fact that there is no apparent dependence
of the mean charges on the coronal distance (e.g., for the flares of
1992 Oct 30 and 1992 Nov 2) may be problematic for either the distributed
acceleration model or the photoionization model.

\acknowledgments

The author would like to thank Paul Evenson, Wolfgang Dr\"oge, 
Valery Ostryakov, and Roman Hat\-zky 
for helpful discussions and for sending important
journal articles.  I am also grateful to Wiwat Youngdee for data entry
assistance.  This work was partially supported by a grant from 
the Thailand Research Fund.


\clearpage

\begin{thebibliography}{}
\bibitem[Arnaud \& Rothenflug 1985]{ar85} Arnaud, M., \& Rothenflug,
R. 1985, A\&AS, 60, 425
\bibitem[Boberg, Tylka, \& Adams 1996]{bta96} Boberg, P. R., Tylka,
J. A., \& Adams, J. H., Jr. 1996, ApJ, 471, L65
\bibitem[Cane, McGuire, \& von Rosenvinge 1986]{cea86} Cane, H. V., 
McGuire, R. E., \& von Rosenvinge, T. T. 1986, \apj, 301, 448
\bibitem[Fisk \& Schatten 1972]{fs72} Fisk, L. A., \& Schatten, K. H.
1972, Solar Phys., 23, 204
\bibitem[Forbush 1946]{f46} Forbush, S. E. 1946, Phys. Rev., 70, 771
\bibitem[Foukal 1990]{f90} Foukal, P. 1990, Solar Astrophysics (New
York: Wiley), p. 401
\bibitem[Heras et al.\ 1995]{hea95} Heras, A. M., Sanahuja, B.,
Lario, D., Smith, Z. K., Detman, T., \& Dryer, M. 1995, \apj, 445,
497
\bibitem[Hovestadt et al.\ 1984]{hea84} Hovestadt, D., Gloeckler, G.,
Klecker, B., \& Scholer, M. 1984, \apj, 281, 463
\bibitem[Kahler, Reames, \& Sheeley 1990]{krs90} Kahler, S., Reames,
D. V., \& Sheeley, N. R., Jr. 1990, Proc. 21st Internat. Cosmic Ray
Conf. (Adelaide), 5, 183
\bibitem[Lee \& Ryan 1986]{lr86} Lee, M. A., \& Ryan, J. M. 1986,
\apj, 303, 829
\bibitem[Leske et al.\ 1995]{lea95} Leske, R. A., Cummings, J. R.,
Mewaldt, R. A., Stone, E. C., and von Rosenvinge, T. T. 1995, \apj,
452, L149
\bibitem[Luhn et al.\ 1984]{lea84} Luhn, A., Klecker, B., Hovestadt,
D., Gloeckler, G., Ipavich, F. M., Scholer, M., Fan, C. Y., \& Fisk,
L. A. 1984, Adv. Space Res., 4, 161
\bibitem[Luhn et al.\ 1987]{lea87} Luhn, A., Klecker, B., Hovestadt,
D., \& M\"obius, E. 1987, \apj, 317, 951
\bibitem[Mason, Gloeckler, \& Hovestadt 1984]{mgh84} Mason, G. M.,
Gloeckler, G., \& Hovestadt, D. 1984, \apj, 280, 902
\bibitem[Mason et al.\ 1995]{mea95} Mason, G. M., Mazur, J. E.,
Looper, M. D., \& Mewaldt, R. A. 1995, \apj, 452, 901
\bibitem[Mazur et al.\ 1992]{mea92} Mazur, J. E., Mason, G. M.,
Klecker, B., \& McGuire, R. E. 1992, \apj, 401, 398
\bibitem[Miller \& Vi\~nas 1993]{mv93} Miller, J. A., \& Vi\~nas, A.
F. 1993, \apj, 412, 386
\bibitem[Mullan \& Waldron 1986]{mw86} Mullan, D. J., \& Waldron, W.
L. 1986, \apj, 308, L21
\bibitem[Newkirk \& Wentzel 1978]{nw78} Newkirk, G., Jr., \& Wentzel,
D. G. 1978, \jgr, 83, 2009
\bibitem[Oetliker et al.\ 1995]{oea95} Oetliker, M., Klecker, B.,
Hovestadt, D., Scholer, M., Blake, J. B., Looper, M., \& Mewaldt, R.
A. 1995, Proc.\ 24th Internat. Cosmic Ray Conf. (Rome), 4, 470
\bibitem[Oetliker et al.\ 1997]{oea97} Oetliker, M., Klecker, B.,
Hovestadt, D., Mason, G. M., Mazur, J. E., Leske, R. A., Mewaldt, R.
A., Blake, J. B., \& Looper, M. D. 1997, ApJ (in press)
\bibitem[Pallavicini et al.\ 1977]{psv77} Pallavicini, R.,
Serio, S., \& Vaiana, G. S. 1977, \apj, 216, 108
\bibitem[Palmer 1982]{p82} Palmer, I. D. 1982, Rev. Geophys. Space
Phys., 20, 335
\bibitem[Reames 1990]{r90} Reames, D. V. 1990, \apj, 358, L63
\bibitem[Reames, Cane, \& von Rosenvinge 1990]{rea90} 
Reames, D. V., Cane, H. V., \&
von Rosenvinge, T. T. 1990, \apj, 357, 259
\bibitem[Reid 1964]{r64} Reid, G. C. 1964, \jgr, 67, 2639
\bibitem[Schwarz et al.\ 1988]{sea88} Schwarz, S. J., Thomsen, M. F.,
Bame, S. J., and Stansberry, J. 1988, \jgr, 93, 12923
\bibitem[Shull \& Van Steenberg 1982]{svs82} Shull, J. M., \& Van
Steenberg, M. 1982, \apjs, 48, 95
\bibitem[Solar-Geophysical Data 1993]{sgd93} Solar-Geophysical Data
1993, (Boulder: US Dept. of Commerce)
\bibitem[Tan et al. 1990]{tea90} Tan, L. C., Mason, G. M., Klecker,
B., \& Hovestadt, D. 1990, 
Proc.\ 21th Internat. Cosmic Ray Conf. (Adelaide), 5, 361
\bibitem[Temerin \& Roth 1992]{tr92} Temerin, M., \& Roth, I. 1992,
\apj, 391, L105
\bibitem[Terasawa \& Scholer 1989]{ts89} Terasawa, T., \& Scholer, M.
1989, Science, 244, 1050
\bibitem[Thomsen et al.\ 1987]{tea87} Thomsen, M. F., Mellott, M. M.,
Stansberry, J. A., Bame, S. J., Gosling, J. T., \& Russell, C. T.
1987, \jgr, 92, 10119
\bibitem[Tylka et al.\ 1995]{tea95} Tylka, A. J., Boberg, P. R.,
Adams, J. H., Jr., Beahm, L. P., Dietrich, W. F., \& Kleis, T. 1995,
\apj, 444, L109
%
\end{thebibliography}
\end{document}